\documentclass[12pt]{article}
\usepackage{epsfig}
\usepackage{amsmath}

\def\beq{\begin{equation}}
\def\eeq#1{\label{#1}\end{equation}}
\def\eeqn{\end{equation}}

\def\beqa{\begin{eqnarray}}
\def\eeqa#1{\label{#1}\end{eqnarray}}
\def\eeqan{\end{eqnarray}}

\let\bar=\overbar

\def\half{\frac{1}{2}}

\def\Dslash{\not{\hbox{\kern-4pt $D$}}}
\def\dslash{\not{\hbox{\kern-2pt $\del$}}}

\def\msb{{\bar{\ssstyle M \kern -1pt S}}}

\def\Title#1{\begin{center} {\Large {\bf #1} } \end{center}}

\newcommand{\mbf}[1]{\mathbf{#1}}

\begin{document}

\Title{Light-Front Holography and AdS/QCD: \\ \vspace{5pt} A First Approximation to QCD}

\bigskip\bigskip


\begin{raggedright}

{\it  Stanley J. Brodsky\\
SLAC National Accelerator Laboratory\\
Stanford University, Stanford, CA 94309, USA\\
\vspace{2pt}
and\\
\vspace{2pt}
Guy F. de T\'eramond\\
Universidad de Costa Rica\\
San Jos\'e, Costa Rica}
\bigskip\bigskip
\end{raggedright}

\vspace{10pt}

\begin{abstract}
The combination of Anti-de Sitter  space (AdS) methods with light-front holography leads to a semi-classical first approximation to the spectrum and wavefunctions of meson and baryon light-quark  bound states.
Starting from the bound-state Hamiltonian equation of motion in QCD, we derive  relativistic light-front wave equations in terms of an invariant impact variable $\zeta$ which measures the separation of the quark and gluonic constituents within the hadron at equal light-front time. These equations of motion in physical space-time are  equivalent to the equations of motion which describe the propagation of spin-$J$ modes in anti--de Sitter (AdS) space. Its eigenvalues give the hadronic spectrum, and its eigenmodes represent the probability distribution of the hadronic constituents at a given scale. Applications to the light meson and baryon spectra are presented. The predicted  meson spectrum has a string-theory Regge form ${\cal M}^2 = 4 \kappa^2(n+L+S/2 )$; {\it i.e.}, the square of the eigenmass is linear in both $L$ and $n$, where $n$ counts the number of nodes  of the wavefunction in the radial variable $\zeta$. The space-like pion form factor is also well reproduced. 
One thus obtains a remarkable
connection between the description of hadronic modes in AdS space and
the Hamiltonian formulation of QCD in physical space-time quantized
on the light-front  at fixed light-front time $\tau.$  The model 
can be systematically improved  by using its complete orthonormal solutions to diagonalize the full QCD light-front Hamiltonian or by applying the Lippmann-Schwinger method in order to systematically include the QCD interaction terms.

\vspace{1pc}
\end{abstract}


\section{Introduction}

The Schr\"odinger equation plays a central role in atomic physics, providing a simple,  but effective, first approximation description of  the spectrum and wavefunctions of bound states in quantum electrodynamics. It can be systematically improved in QED, leading to an exact quantum field theoretic description of  atomic states such as positronium and muonium as given by the  relativistic Bethe-Salpeter equation.

We have recently shown that  the combination of Anti-de Sitter  space (AdS) methods with light-front holography lead to a remarkably accurate first approximation for the spectrum and wavefunctions of meson and baryon light-quark  bound states. The resulting equation for a meson $q \bar q$ bound state at fixed light-front time $\tau= t +z/c$ , the time marked by the
front of a light wave,~\cite{Dirac:1949cp}  has the form of a relativistic Lorentz invariant  Schr\"odinger equation
\begin{equation} \label{eq:QCDLFWE}
\left(-\frac{d^2}{d\zeta^2}
- \frac{1 - 4L^2}{4\zeta^2} + U(\zeta) \right)
\phi(\zeta) = \mathcal{M}^2 \phi(\zeta),
\end{equation}
where the confining potential is $ U(\zeta) = \kappa^4 \zeta^2 + 2 \kappa^2(L+S-1)$ 
in a soft dilaton modified background.
There is only one parameter, the mass scale $\kappa \sim 1/2$ GeV, which enters the confinement potential. Here $S=0,1$ is the spin of the $q $ and $ \bar q $, $L$ is their relative orbital angular momentum and $\zeta = \sqrt{x(1-x)b^2_\perp}$ is a Lorentz invariant coordinate that measures 
the distance between the quark and antiquark; it is analogous to the radial coordinate $r$ in the Schrodinger equation.
One thus obtains a remarkable
connection between the description of hadronic modes in AdS space and
the Hamiltonian formulation of QCD in physical space-time quantized
on the light-front (LF) at fixed light-front time $\tau.$

A key step in the analysis of an atomic system such as positronium
is the introduction of the spherical coordinates $r, \theta, \phi$
which  separates the dynamics of Coulomb binding from the
kinematical effects of the quantized orbital angular momentum $L$.
The essential dynamics of the atom is specified by the radial
Schr\"odinger equation whose eigensolutions $\psi_{n,L}(r)$
determine the bound-state wavefunction and eigenspectrum. In our recent
work,~\cite{deTeramond:2008ht} we have shown that there is an analogous invariant
light-front coordinate $\zeta$ which allows one to separate the
essential dynamics of quark and gluon binding from the kinematical
physics of constituent spin and internal orbital angular momentum.
The result is the single-variable LF Schr\"odinger equation for QCD  Eq.  \ref{eq:QCDLFWE},
which determines the eigenspectrum and the light-front wavefunctions (LFWFs)
of hadrons for general spin and orbital angular momentum.~\cite{deTeramond:2008ht}  If one further chooses  the constituent rest frame (CRF)~\cite{Danielewicz:1978mk,Karmanov:1979if,Glazek:1983ba}  
where the total 3-momentum vanishes: $\sum^n_{i=1} \mbf{k}_i \! = \! 0$, then the kinetic energy in the LF wave equation displays the usual 3-dimensional rotational invariance. Note that if the binding energy is nonzero, $P^z \ne 0,$ in this frame.

\begin{figure}[!]
\begin{center}
\includegraphics[angle=0,width=8.5cm]{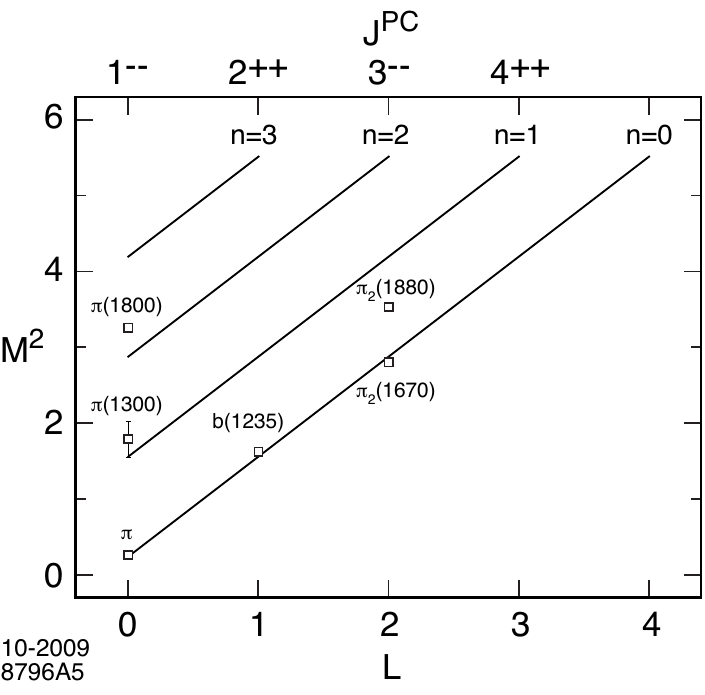}
\caption{Parent and daughter Regge trajectories for the $\pi$-meson family for  
$\kappa= 0.6$ GeV.}
\label{pion}
\end{center}
\end{figure}

\begin{figure}[!]
\begin{center}
\includegraphics[angle=0,width=8.5cm]{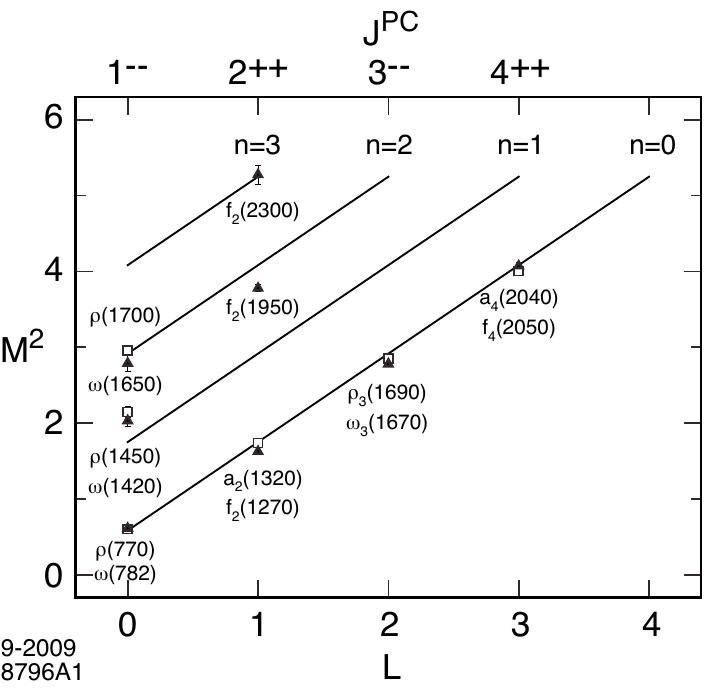}
\caption{Regge trajectories for the  $I\!=\!1$ $\rho$-meson
 and the $I\!=\!0$  $\omega$-meson families for $\kappa= 0.54$ GeV.}
\label{VM}
\end{center}
\end{figure}

The meson spectrum predicted by  Eq. \ref{eq:QCDLFWE} has a string-theory Regge form
${\cal M}^2 = 4 \kappa^2(n+ L+S/2)$; {\it i.e.}, the square of the eigenmasses are linear in both $L$ and $n$, where $n$ counts the number of nodes  of the wavefunction in the radial variable $\zeta$.  This is illustrated for the pseudoscalar and vector meson spectra in Figs. \ref{pion} and \ref{VM},
where the data are from Ref.  \cite{Amsler:2008xx}.
The pion ($S=0, n=0, L=0$) is massless for zero quark mass, consistent with chiral invariance.  Thus one can compute the hadron spectrum by simply adding  $4 \kappa^2$ for a unit change in the radial quantum number, $4 \kappa^2$ for a change in one unit in the orbital quantum number  and $2 \kappa^2$ for a change of one unit of spin $S$. Remarkably, the same rule holds for three-quark baryons as we shall show below. For other 
recent calculations of the hadronic spectrum based on AdS/QCD, see Refs.~\cite{Boschi-Filho:2002vd,   BoschiFilho:2005yh, Evans:2006ea, Hong:2006ta, Colangelo:2007pt, Forkel:2007ru, Vega:2008af, Nawa:2008xr, dePaula:2008fp,  Colangelo:2008us, Forkel:2008un, Ahn:2009px, dePaula:2009za,Sui:2009xe}.

\begin{figure}[!]
\begin{center}
\includegraphics[width=9.5cm]{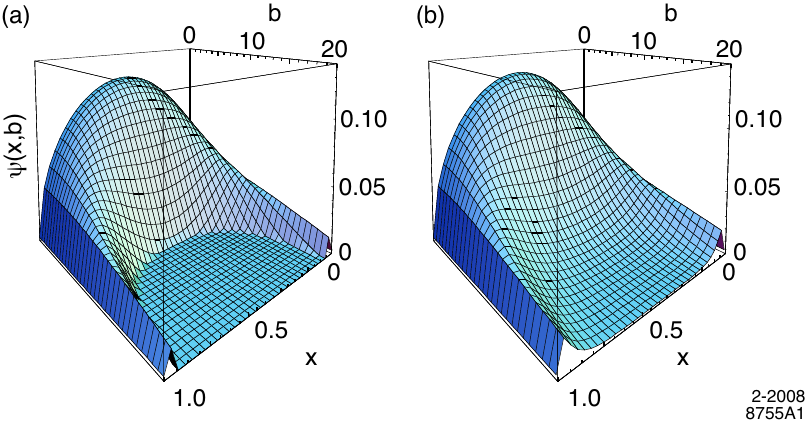}
 \caption{Pion light-front wavefunction $\psi_\pi(x, \mbf{b}_\perp$) for the  AdS/QCD (a) hard-wall ($\Lambda_{QCD} = 0.32$ GeV) and (b) soft-wall  ( $\kappa = 0.375$ GeV)  models.}
\label{LFWF}
\end{center}
\end{figure}

\begin{figure}[h]
\centering
\includegraphics[angle=0,width=8.0cm]{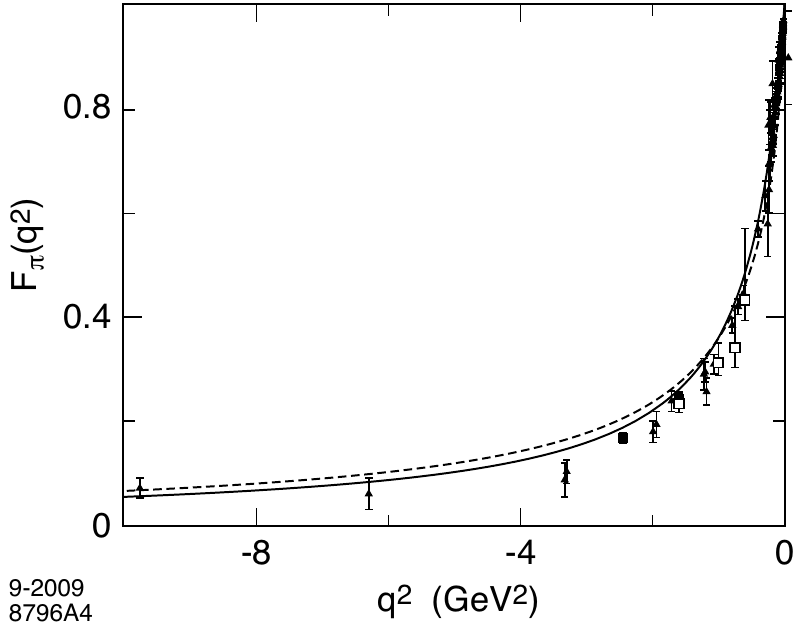}
\caption{Space-like scaling behavior for $F_\pi(Q^2)$ as a function of $q^2$.
The continuous line is the prediction of the soft-wall model for  $\kappa = 0.375$ GeV.
The dashed line is the prediction of the hard-wall model for $\Lambda_{\rm QCD} = 0.22$ GeV.
The triangles are the data compilation  from Baldini  {\it et al.},~\cite{Baldini:1998qn}  the  filled boxes  are JLAB 1 data~\cite{Tadevosyan:2007yd} and empty boxes are JLAB 2
 data.~\cite{Horn:2006tm}}
\label{PionFFSL}
\end{figure}

The eigensolutions of  Eq. \ref{eq:QCDLFWE} provide the light-front wavefunctions of the valence Fock state of the hadrons $
\psi(\zeta) = \psi(x, \mbf{b}_\perp)$
 as illustrated for the pion in Fig. \ref{LFWF} for the soft and hard wall models.  Given these wavefunctions one can predict many hadronic observables such as the generalized parton distributions that enter deeply virtual Compton scattering. For example, hadron form factors can be predicted from the overlap of LFWFs as in the Drell-Yan West formula.
The prediction for the space-like pion form factor is shown in Fig.  \ref{PionFFSL}.
The pion form factor and the vector meson poles residing in the dressed current in the soft wall model require choosing  a value of $\kappa$ smaller by a factor of $1/\sqrt 2$  than the canonical value of  $\kappa$ which determines the mass scale of the hadronic spectra.  This shift is apparently due to the fact that the transverse current in $e^+ e^- \to q \bar q$ creates a quark pair with $L^z= \pm 1$ instead of the $L^z=0$ $q \bar q$ composition of the vector mesons in the spectrum. We will discuss this further in an upcoming paper.
Other recent computations of the space-like pion form factor in AdS/QCD are presented in~\cite{Kwee:2007dd,Grigoryan:2007wn}. Given
the LFWFs one can compute jet hadronization at the amplitude level from first principles.~\cite{Brodsky:2008tk} A similar method has been used to predict the production of antihydrogen from the off-shell coalescence of relativistic antiprotons and positrons.~\cite{Munger:1993kq}

This semi-classical first approximation to QCD can be systematically improved  by using the complete orthonormal solutions of Eq. \ref{eq:QCDLFWE} to diagonalize the QCD light-front Hamiltonian~\cite{Vary:2009gt}  or by applying the Lippmann-Schwinger method to systematically include the QCD interaction terms. In either case, the result is the full Fock state structure of the hadron eigensolution. One can also model 
heavy-light
and heavy hadrons by including non-zero quark masses in the LF kinetic energy 
$\sum_i ({\mbf{k}^2_{\perp  i}+ m^2_i)/x_i}$
 as well as the effects of the one-gluon exchange potential.

One can derive these results  in two parallel ways. In the first method, one begins with a conformal approximation to QCD, justified by evidence that the theory has an infrared fixed point.~\cite{Deur:2008rf}  One then uses the fact that the conformal group has a geometrical representation in the five-dimensional AdS$_5$ space
to model an effective dual gravity description in AdS. The fact that
conformal invariance is reflected in the isometries of AdS is an essential ingredient of Maldacena's AdS/CFT correspondence.  Confinement is
effectively introduced with a sharp cut-off in the infrared region of AdS space, the ``hard-wall" model,~\cite{Polchinski:2001tt}
 or with a dilaton background in the fifth dimension which produces a smooth cutoff and linear Regge trajectories,
the ``soft-wall" model.~\cite{Karch:2006pv}
The soft-wall AdS/CFT model with a dilaton-modified AdS space leads to the
 potential $U(z) = \kappa^4 z^2 + 2 \kappa^2(L+S-1)$. This potential can be derived directly from the action in AdS 
 space~\cite{deTeramond:2009xx}
 and corresponds to a dilaton profile $\exp(+\kappa^2 z^2)$, with  opposite sign  to that of Ref.  \cite{Karch:2006pv}.  
 The introduction of a positive dilaton profile is also relevant for describing chiral symmetry breaking,~\cite{Zuo:2009dz}
 since  the expectation value of the scalar field associated with the quark mass and condensate does not blow up in the far infrared region of AdS in contrast with the original model.~\cite{Karch:2006pv}
 Glazek and Schaden~\cite{Glazek:1987ic} have shown that a  harmonic oscillator confining potential naturally arises as an effective potential between heavy quark states when one stochastically eliminates higher gluonic Fock states. Also, Hoyer~\cite{Hoyer:2009ep} has argued that the Coulomb  and  a linear  potentials are uniquely allowed in the Dirac equation at the classical level. The linear potential  becomes a harmonic oscillator potential in the corresponding Klein-Gordon equation.

 Hadrons are identified by matching the power behavior of the hadronic amplitude at the AdS boundary at small $z$ to the twist of its interpolating operator at short distances $x^2 \to 0$, as required by the AdS/CFT dictionary. The twist corresponds to the dimension of fields appearing in chiral super-multiplets.~\cite{Craig:2009rk}
The twist of a hadron equals the number of constituents.
We  then apply light-front holography to relate the amplitude eigensolutions  in the fifth dimension coordinate $z$  to the LF wavefunctions in the physical spacetime variable  $\zeta.$  Light-front holography can be derived by establishing an identity between the Polchinski-Strassler formula for current matrix elements and the corresponding Drell-Yan-West
formula in LF theory.  The same correspondence is obtained for both electromagnetic and gravitational form factors, a nontrivial test of consistency.

In the second method we  use a first semiclassical approximation  to transform the fixed
LF time bound-state Hamiltonian equation  to  a corresponding wave equation in AdS space.
The  invariant LF
coordinate $\zeta$ allows the separation of the dynamics of quark and gluon binding from
the kinematics of constituent spin and internal orbital angular momentum.~\cite{deTeramond:2008ht}  In effect $\zeta$ represents the off-light-front energy shell or invariant mass dependence of the bound state.
The result is the  single-variable LF relativistic
Schr\"odinger equation  which determines the spectrum
and LFWFs  of hadrons for general spin and
orbital angular momentum. This LF wave equation serves as a semiclassical first
approximation to QCD, and it is equivalent to the
equations of motion which describe the propagation of spin-$J$ modes
in  AdS space.

The term $-{(1\! - 4L^2)/ 4 \zeta^2}$  in the  LF equation of motion  (\ref{eq:QCDLFWE})
is derived from  the reduction of the LF kinetic energy when one transforms to the radial $\zeta$  and angular coordinate 
$\varphi$, in analogy to the $\ell(\ell+1)/ r^2$ Casimir term in Schr\"odinger theory.  One thus establishes the interpretation of $L$ in the AdS equations of motion.
The interaction terms build confinement and correspond to
truncation of AdS space~\cite{deTeramond:2008ht} in an effective dual gravity  approximation.
The duality between these two methods provides a direct
connection between the description of hadronic modes in AdS space and
the Hamiltonian formulation of QCD in physical space-time quantized
on the light-front  at fixed LF time $\tau.$

The identification of orbital angular momentum of the constituents is a key element in the description of the internal structure of hadrons using holographic principles. In our approach  quark and gluon degrees of freedom are explicitly introduced in the gauge/gravity correspondence, in contrast with the usual
AdS/QCD framework~\cite{Erlich:2005qh,DaRold:2005zs} where axial and vector currents become the primary entities as in effective chiral theory. 
In our approach
the holographic mapping is carried out in the  strongly coupled regime where QCD is almost conformal corresponding to an infrared fixed-point.~\cite{Deur:2008rf}
Our analysis follows from recent developments in light-front
QCD~\cite{Brodsky:2003px,deTeramond:2005su,Brodsky:2006uqa,Brodsky:2007hb,Brodsky:2008pf} which have been inspired by the
AdS/CFT correspondence~\cite{Maldacena:1997re}.

In the standard applications of  AdS/CFT methods, one begins with Maldacena's duality between  the conformal supersymmetric $SO(4,2)$
 gauge theory and a semiclassical supergravity string theory defined in a 10 dimension 
 AdS$_5 \times S^5$
 space-time.  There are no existing
 string theories actually dual to QCD, but nevertheless many interesting predictions can be made. In contrast, in our approach, we simply use the mathematical fact that the effects of scale transformations in a conformal theory can be mapped to the $z$ dependence of amplitudes in AdS$_5$ space. QCD has an infrared fixed point and a conformal window in the infrared domain since the propagators of the confined quarks and gluons in the  loop integrals contributing to the $\beta$ function have a maximal wavelength.~\cite{Brodsky:2008be} Just as in QED, the $\beta$ function  must vanish in the IR. One then uses AdS$_5$ to represent scale transformations within the conformal window. Unlike the top-down supergravity approach,  one is not limited to hadrons of maximum spin 
$J \le 2$ in our bottom-up approach, and one can study baryons with $N_C=3.$

\section{The Light-Front Hamiltonian Approach to QCD}

The
Schr\"odinger wavefunction describes the quantum-mechanical
structure of  an atomic system at the amplitude level. Light-front
wavefunctions play a similar role in quantum chromodynamics,
providing a fundamental description of the structure and
internal dynamics of hadrons in terms of their constituent quarks
and gluons. The LFWFs of bound states in QCD are
relativistic generalizations of the Schr\"odinger wavefunctions of
atomic physics, but they are determined at fixed light-cone time
$\tau  = t +z/c$ -- the ``front form'' introduced by
Dirac~\cite{Dirac:1949cp} -- rather than at fixed ordinary time $t.$

When a flash from a camera illuminates a scene, each object is illuminated along the light-front of the flash; i.e., at a given $\tau$.  Similarly, when a sample is illuminated by an x-ray source, each element of the target is struck at a given $\tau.$  In contrast, setting the initial condition using conventional instant time $t$ requires simultaneous scattering of photons on each constituent.
Thus it is natural to set boundary conditions at fixed $\tau$ and then evolve the system using the light-front (LF) Hamiltonian
$P^-  \!= \!  P^0-P^3 = i {d/d \tau}.$  The invariant Hamiltonian $H_{LF} = P^+ P^- \! - P^2_\perp$ then has eigenvalues $\mathcal{M}^2$ where $\mathcal{M}$ is the physical mass.   Its eigenfunctions are the light-front eigenstates whose Fock state projections define the light-front wavefunctions.  Given the LF Fock state wavefunctions
$\psi^H_n(x_i, \mbf{k}_{\perp i}, \lambda_i),$
where $x_i \! =\! k^+/P^+$,  $\sum_{i=1}^n x_i  \! = \! 1, ~ \sum_{i=1}^n \mbf{k}_{\perp i}  \! = \! 0$, one
can immediately compute observables such as hadronic form factors (overlaps of LFWFs), structure
functions (squares of LFWFS), as well as the generalized parton distributions and
distribution amplitudes which underly hard exclusive reactions.

The most useful  feature of LFWFs is the fact that they are frame
independent; i.e., the form of the LFWF is independent of the
hadron's total momentum $P^+ = P^0 + P^3$ and $\mbf{P}_\perp.$
The simplicity of Lorentz boosts of LFWFs contrasts dramatically with the complexity of the boost of wavefunctions defined at fixed time $t.$~\cite{Brodsky:1968ea}
Light-front quantization is thus the ideal framework to describe the
structure of hadrons in terms of their quark and gluon degrees of freedom.  The
constituent spin and orbital angular momentum properties of the
hadrons are also encoded in the LFWFs.
The total  angular momentum projection~\cite{Brodsky:2000ii},
$J^z = \sum_{i=1}^n  S^z_i + \sum_{i=1}^{n-1} L^z_i$,
is conserved Fock-state by Fock-state and by every interaction in the LF Hamiltonian.
Other advantageous features of light-front quantization include:

\begin{itemize}

\item
The simple structure of the light-front vacuum allows an unambiguous
definition of the partonic content of a hadron in QCD.  The chiral and gluonic condensates are properties of the higher Fock states,~\cite{Casher:1974xd,Brodsky:2009zd} rather than the vacuum.  In the case of the Higgs model, the effect of the usual Higgs vacuum expectation value is replaced by a constant $k^+=0$ zero mode field.~\cite{Srivastava:2002mw}

\item
If one quantizes QCD in the physical light-cone gauge (LCG) $A^+ =0$, then gluons only have physical angular momentum projections $S^z= \pm 1$. The orbital angular momenta of quarks and gluons are defined unambiguously, and there are no ghosts.

\item
The gauge-invariant distribution amplitude $\phi(x,Q)$  is the integral of the valence LFWF in LCG integrated over the internal transverse momentum $k^2_\perp < Q^2$ because the Wilson line is trivial in this gauge. It is also possible to quantize QCD in  Feynman gauge in the light front.~\cite{Srivastava:1999gi}

\item
LF Hamiltonian perturbation theory provides a simple method for deriving analytic forms for the analog of Parke-Taylor amplitudes~\cite{Motyka:2009gi} where each particle spin $S^z$ is quantized in the LF $z$ direction.  The gluonic $g^6$ amplitude  $T(-1 -1 \to +1 +1 +1 +1 +1 +1)$  requires $\Delta L^z =8;$ it thus must vanish at tree level since each three-gluon vertex has  $\Delta L^z = \pm 1.$ However, the order $g^8$ one-loop amplitude can be nonzero.

\item
Amplitudes in light-front perturbation theory are automatically renormalized using the ``alternate denominator"  subtraction method.~\cite{Brodsky:1973kb} The application to QED has been checked at one and two loops.~\cite{Brodsky:1973kb}

\item
One can easily show using LF quantization that the anomalous gravitomagnetic moment $B(0)$  of a nucleon, as  defined from the spin flip matrix element of the gravitational current, vanishes Fock-state by Fock state~\cite{Brodsky:2000ii}, as required by the equivalence principle.~\cite{Teryaev:1999su}

\item
LFWFs obey the cluster decomposition theorem, providing an elegant proof of this theorem for relativistic bound states.~\cite{Brodsky:1985gs}

\item
The LF Hamiltonian can be diagonalized using the discretized light-cone quantization (DLCQ) method.~\cite{Pauli:1985ps} This nonperturbative method is particularly useful for solving low-dimension quantum field theories such as
QCD$(1+1).$~\cite{Hornbostel:1988fb}

\item
LF quantization provides a distinction between static  (the square of LFWFs) distributions versus non-universal dynamic structure functions,  such as the Sivers single-spin correlation and diffractive deep inelastic scattering which involve final state interactions.  The origin of nuclear shadowing and process independent anti-shadowing also becomes explicit.

\item
LF quantization provides a simple method to implement jet hadronization at the amplitude level.

\item
The instantaneous fermion interaction in LF  quantization provides a simple derivation of the $J=0$
fixed pole contribution to deeply virtual Compton scattering,~\cite{Brodsky:2009bp} i.e., the $e^2_q s^0 F(t)$  contribution to the DVCS amplitude which is independent of photon energy and virtuality.

\item
Unlike instant time quantization, the bound state
Hamiltonian equation of motion in the LF is frame independent. This makes a direct connection of QCD with AdS/CFT methods possible.~\cite{deTeramond:2008ht}

\end{itemize}

\section{The Light-Front Hamiltonian}

Light-front quantization of QCD provides a nonperturbative method for  solving  QCD in Minkowski space. Unlike lattice gauge theories, fermions introduce no new complications.

We can  define the  LF Lorentz invariant Hamiltonian
$H_{LF}= P_\mu P^\mu = P^-P^+  \! - \mbf{P}^2_\perp$ with eigenstates
$\vert \psi_H(P^+, \mbf{P}_\perp, S_z )\rangle$
and eigenmass  $\mathcal{M}_H^2$, the mass spectrum of the color-singlet states
of QCD~\cite{Brodsky:1997de}
\begin{equation} \label{eq:HLF}
H_{LF} \vert \psi_H\rangle = {\cal M}^2_H \vert \psi_H \rangle.
\end{equation}
A state $\vert \psi_H \rangle$ is an expansion
in multi-particle Fock states
$\vert n \rangle $ of the free LF Hamiltonian:
~$\vert \psi_H \rangle = \sum_n \psi_{n/H} \vert n \rangle$, where
a one parton state is $\vert q \rangle = \sqrt{2 q^+} \,b^\dagger(q) \vert 0 \rangle$.
The Fock components $\psi_{n/H}(x_i, {\mathbf{k}_{\perp i}}, \lambda_i^z)$
are independent of  $P^+$ and $\mbf{P}_{\! \perp}$
and depend only on relative partonic coordinates:
the momentum fraction
 $x_i = k^+_i/P^+$, the transverse momentum  ${\mathbf{k}_{\perp i}}$ and spin
 component $\lambda_i^z$.
 The LFWFs $\psi_{n/H}$ provide a
{\it frame-independent } representation of a hadron which relates its quark
and gluon degrees of freedom to their asymptotic hadronic state.

One can derive light-front holography using a first semiclassical approximation  to transform the fixed
light-front time bound-state Hamiltonian equation of motion in QCD  to  a corresponding wave equation in AdS
space.~\cite{deTeramond:2008ht} To this end we
 compute the invariant hadronic mass $\mathcal{M}^2$ from the hadronic matrix element
\begin{equation}
\langle \psi_H(P') \vert H_{LF}\vert\psi_H(P) \rangle  =
\mathcal{M}_H^2  \langle \psi_H(P' ) \vert\psi_H(P) \rangle,
\end{equation}
expanding the initial and final hadronic states in terms of its Fock components. We use the
frame $P = \big(P^+, M^2/P^+, \vec{0}_\perp \big)$ where $H_{LF} =  P^+ P^-$.
The LF expression for $\mathcal{M}^2$ In impact space is
\begin{multline}
 \mathcal{M}_H^2  =  \sum_n  \prod_{j=1}^{n-1} \int d x_j \, d^2 \mbf{b}_{\perp j} \,
\psi_n^*(x_j, \mbf{b}_{\perp j})    \,
 \sum_q   \left(\frac{ \mbf{- \nabla}_{ \mbf{b}_{\perp q}}^2  \! + m_q^2 }{x_q} \right)
 \psi_n(x_j, \mbf{b}_{\perp j}) \\
  + {\rm (interactions)} , \label{eq:Mb}
 \end{multline}
plus similar terms for antiquarks and gluons ($m_g = 0)$.

To simplify the discussion we will consider a two-parton hadronic bound state.  In the limit
of zero quark mass
$m_q \to 0$
\begin{equation}  \label{eq:MKb}
\mathcal{M}^2  =  \int_0^1 \! \frac{d x}{x(1-x)} \int  \! d^2 \mbf{b}_\perp  \,
  \psi^*(x, \mbf{b}_\perp)
  \left( - \mbf{\nabla}_{ {\mbf{b}}_{\perp}}^2\right)
  \psi(x, \mbf{b}_\perp) +   {\rm (interactions)}.
 \end{equation}
 The functional dependence  for a given Fock state is
given in terms of the invariant mass
\begin{equation}
 \mathcal{M}_n^2  = \Big( \sum_{a=1}^n k_a^\mu\Big)^2 = \sum_a \frac{\mbf{k}_{\perp a}^2 +  m_a^2}{x_a}
 \to \frac{\mbf{k}_\perp^2}{x(1-x)} \,,
 \end{equation}
 the measure of the off-mass shell energy~ $\mathcal{M}^2 - \mathcal{M}_n^2$
 of the bound state.
 Similarly in impact space the relevant variable for a two-parton state is  $\zeta^2= x(1-x)\mbf{b}_\perp^2$.
Thus, to first approximation  LF dynamics  depend only on the boost invariant variable
$\mathcal{M}_n$ or $\zeta,$
and hadronic properties are encoded in the hadronic mode $\phi(\zeta)$ from the relation
\begin{equation} \label{eq:psiphi}
\psi(x,\zeta, \varphi) = e^{i M \varphi} X(x) \frac{\phi(\zeta)}{\sqrt{2 \pi \zeta}} ,
\end{equation}
thus factoring out the angular dependence $\varphi$ and the longitudinal, $X(x)$, and transverse mode $\phi(\zeta)$
with normalization $ \langle\phi\vert\phi\rangle = \int \! d \zeta \,
 \vert \langle \zeta \vert \phi\rangle\vert^2 = 1$.

We can write the Laplacian operator in (\ref{eq:MKb}) in circular cylindrical coordinates $(\zeta, \varphi)$
and factor out the angular dependence of the
modes in terms of the $SO(2)$ Casimir representation $L^2$ of orbital angular momentum in the
transverse plane. Using  (\ref{eq:psiphi}) we find~\cite{deTeramond:2008ht}
\begin{equation} \label{eq:KV}
\mathcal{M}^2   =  \int \! d\zeta \, \phi^*(\zeta) \sqrt{\zeta}
\left( -\frac{d^2}{d\zeta^2} -\frac{1}{\zeta} \frac{d}{d\zeta}
+ \frac{L^2}{\zeta^2}\right)
\frac{\phi(\zeta)}{\sqrt{\zeta}}
+ \int \! d\zeta \, \phi^*(\zeta) U(\zeta) \phi(\zeta) ,
\end{equation}
where all the complexity of the interaction terms in the QCD Lagrangian is summed up in the effective potential $U(\zeta)$.
The LF eigenvalue equation $H_{LF} \vert \phi \rangle  =  \mathcal{M}^2 \vert \phi \rangle$
is thus a light-front  wave equation for $\phi$
\begin{equation} \label{eq:QCDLFWE2}
\left(-\frac{d^2}{d\zeta^2}
- \frac{1 - 4L^2}{4\zeta^2} + U(\zeta) \right)
\phi(\zeta) = \mathcal{M}^2 \phi(\zeta),
\end{equation}
an effective single-variable light-front Schr\"odinger equation which is
relativistic, covariant and analytically tractable. 
It is important to notice that in the light-front  the $SO(2)$ Casimir for orbital angular momentum $L^2$
is a kinematical quantity, in contrast with the usual $SO(3)$ Casimir $\ell(\ell+1)$ from non-relativistic physics which is
rotational, but not boost invariant. Using (\ref{eq:Mb}) one can readily
generalize the equations to allow for the kinetic energy of massive
quarks.~\cite{Brodsky:2008pg}  In this case, however,
the longitudinal mode $X(x)$ does not decouple from the effective LF bound-state equations.

As the simplest example, we consider a bag-like model
where  partons are free inside the hadron
and the interaction terms effectively build confinement. The effective potential is a hard wall:
$U(\zeta) = 0$ if  $\zeta \le 1/\Lambda_{\rm QCD}$ and
 $U(\zeta) = \infty$ if $\zeta > 1/\Lambda_{\rm QCD}$,
 where boundary conditions are imposed on the
 boost invariant variable $\zeta$ at fixed light-front time.   If $L^2 \ge 0$ the LF Hamiltonian is positive definite
 $\langle \phi \vert H_{LF} \vert \phi \rangle \ge 0$ and thus $\mathcal M^2 \ge 0$.
 If $L^2 < 0$ the bound state equation is unbounded from below and the particle
 ``falls towards the center''. The critical value corresponds to $L=0$.
  The mode spectrum  follows from the boundary conditions
 $\phi \! \left(\zeta = 1/\Lambda_{\rm QCD}\right) = 0$, and is given in
 terms of the roots of Bessel functions: $\mathcal{M}_{L,k} = \beta_{L, k} \Lambda_{\rm QCD}$.
 Upon the substitution $\Phi(\zeta) \sim \zeta^{3/2} \phi(\zeta)$, $\zeta \to z$
 we find
 \begin{equation} \label{eq:eomPhiJz}
\left[ z^2 \partial_z^2 - 3 z \, \partial_z + z^2 \mathcal{M}^2
\!  -  (\mu R)^2 \right] \!  \Phi_J  = 0,
\end{equation}
 the wave equation which describes the propagation of a scalar mode in a fixed AdS$_5$ background with AdS radius $R$.
 The five dimensional mass $\mu$ is related to the orbital angular momentum of the hadronic bound state by
 $(\mu R)^2 = - 4 + L^2$. The quantum mechanical stability $L^2 >0$ is thus equivalent to the
 Breitenlohner-Freedman stability bound in AdS.~\cite{Breitenlohner:1982jf}
The scaling dimensions are $\Delta = 2 + L$ independent of $J$ in agreement with the
twist scaling dimension of a two parton bound state in QCD. Higher spin-$J$ wave equations
are obtained by shifting dimensions: $\Phi_J(z) = (z/R)^{-J} \Phi(z)$.~\cite{deTeramond:2008ht}

The hard-wall LF model discussed here is equivalent to the hard wall model of
 Ref.~\cite{Polchinski:2001tt}.   The variable $\zeta$,
 $0 \leq \zeta \leq \Lambda_{\rm QCD}^{-1}$,  represents the invariable separation between pointlike constituents and is also
 the holographic variable $z$ in AdS, thus we can identify $\zeta = z$.
 Likewise a two-dimensional  oscillator with
 effective potential  $ U(z) = \kappa^4 z^2 + 2 \kappa^2(L+S-1)$ is similar to the soft-wall model of
 Ref.~\cite{Karch:2006pv} which reproduce the usual linear Regge trajectories, where $L$ is the internal
 orbital angular momentum and $S$ is the internal spin. The soft-wall discussed here correspond to a positive sign
 dilaton~\cite{deTeramond:2009xx,Zuo:2009dz}, and higher-spin solutions follow from shifting 
 dimensions: $\Phi_J(z) = (z/R)^{-J} \Phi(z)$.

Individual hadron states can be identified by their interpolating operator at $z\to 0.$  For example, the pseudoscalar meson interpolating operator
$\mathcal{O}_{2+L}= \bar q \gamma_5 D_{\{\ell_1} \cdots D_{\ell_m\}} q$,
written in terms of the symmetrized product of covariant
derivatives $D$ with total internal  orbital
momentum $L = \sum_{i=1}^m \ell_i$, is a twist-two, dimension $3 + L$ operator
with scaling behavior determined by its twist-dimension $ 2 + L$. Likewise
the vector-meson operator
$\mathcal{O}_{2+L}^\mu = \bar q \gamma^\mu D_{\{\ell_1} \cdots D_{\ell_m\}} q$
has scaling dimension $\Delta=2 + L$.  The scaling behavior of the scalar and vector AdS modes $\Phi(z) \sim z^\Delta$ at $z \to 0$  is precisely the scaling required to match the scaling dimension of the local pseudoscalar and vector-meson interpolating operators.
The spectral predictions for the light pseudoscalar and
vector mesons  in the Chew-Frautschi plot in Fig. \ref{pion} and \ref{VM} for the soft-wall model discussed here are in good agreement for the principal and daughter Regge trajectories. Radial excitations correspond to the nodes
of the wavefunction.

For baryons, the light-front wave equation is a linear equation
determined by the LF transformation properties of spin 1/2 states. A linear confining potential
$U(\zeta) \sim \kappa^2 \zeta$ in the LF Dirac
equation leads to linear Regge trajectories.~\cite{Brodsky:2008pg}   For fermionic modes the light-front matrix
Hamiltonian eigenvalue equation $D_{LF} \vert \psi \rangle = \mathcal{M} \vert \psi \rangle$, $H_{LF} = D_{LF}^2$,
in a $2 \times 2$ spinor  component
representation is equivalent to the system of coupled linear equations
\begin{eqnarray} \label{eq:LFDirac} \nonumber
- \frac{d}{d\zeta} \psi_- -\frac{\nu+\half}{\zeta}\psi_-
- \kappa^2 \zeta \psi_-&=&
\mathcal{M} \psi_+, \\ \label{eq:cD2k}
  \frac{d}{d\zeta} \psi_+ -\frac{\nu+\half}{\zeta}\psi_+
- \kappa^2 \zeta \psi_+ &=&
\mathcal{M} \psi_-.
\end{eqnarray}
with eigenfunctions
\begin{eqnarray} \nonumber
\psi_+(\zeta) &\sim& z^{\frac{1}{2} + \nu} e^{-\kappa^2 \zeta^2/2}
  L_n^\nu(\kappa^2 \zeta^2) ,\\
\psi_-(\zeta) &\sim&  z^{\frac{3}{2} + \nu} e^{-\kappa^2 \zeta^2/2}
 L_n^{\nu+1}(\kappa^2 \zeta^2),
\end{eqnarray}
and  eigenvalues
\begin{equation}
\mathcal{M}^2 = 4 \kappa^2 (n + \nu + 1) .
\end{equation}

The baryon interpolating operator
$ \mathcal{O}_{3 + L} =  \psi D_{\{\ell_1} \dots
 D_{\ell_q } \psi D_{\ell_{q+1}} \dots
 D_{\ell_m\}} \psi$,  $L = \sum_{i=1}^m \ell_i$ is a twist 3,  dimension $9/2 + L$ with scaling behavior given by its
 twist-dimension $3 + L$. We thus require $\nu = L+1$ to match the short distance scaling behavior. Higher spin fermionic modes are obtained by shifting dimensions for the fields as in the bosonic case.
Thus, as in the meson sector,  the increase  in the 
mass squared for  higher baryonic state is
$\Delta n = 4 \kappa^2$, $\Delta L = 4 \kappa^2$ and $\Delta S = 2 \kappa^2,$ 
relative to the lowest ground state,  the proton.

\begin{figure}[!]
\begin{center}
\includegraphics[angle=0,width=14.0cm]{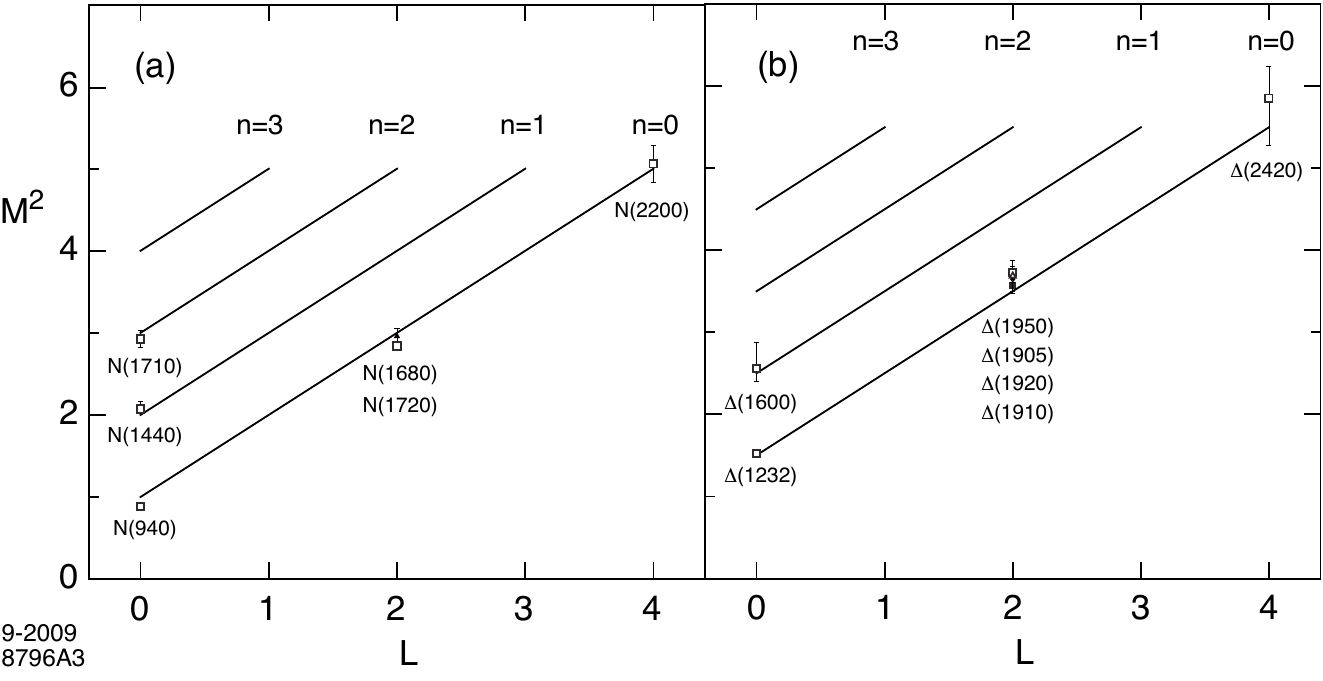}
\caption{{{\bf 56} Regge trajectories for  the  $N$ and $\Delta$ baryon families for $\kappa= 0.5$ GeV}
}
\label{Baryons}
\end{center}
\end{figure}

The predictions for the $\bf 56$-plet of light baryons under the $SU(6)$  flavor group are shown in Fig. \ref{Baryons}.
As for the predictions for mesons in Fig. \ref{VM}, only confirmed PDG~\cite{Amsler:2008xx} states are shown.
The Roper state $N(1440)$ and the $N(1710)$ are well accounted for in this model as the first  and second radial
states. Likewise the $\Delta(1660)$ corresponds to the first radial state of the $\Delta$ family. The model is  successful in explaining the important parity degeneracy observed in the light baryon spectrum, such as the $L\! =\!2$, $N(1680)\!-\!N(1720)$ degenerate pair and the $L=2$, $\Delta(1905), \Delta(1910), \Delta(1920), \Delta(1950)$ states which are degenerate
within error bars. Parity degeneracy of baryons is also a property of the hard wall model, but radial states are not well described in this model.~\cite{deTeramond:2005su}

\section{Holographic Mapping of Transition Amplitudes}

Light-Front Holography can also
be derived by observing the correspondence between matrix elements obtained in AdS/CFT with the corresponding formula using the LF
representation.~\cite{Brodsky:2006uqa}  The light-front electromagnetic form factor in impact
space~\cite{Brodsky:2006uqa,Brodsky:2007hb,Soper:1976jc} can be written as a sum of overlap of light-front wave functions of the $j = 1,2, \cdots, n-1$ spectator
constituents:
\begin{equation} \label{eq:FFb}
F(q^2) =  \sum_n  \prod_{j=1}^{n-1}\int d x_j d^2 \mbf{b}_{\perp j}   \sum_q e_q
            \exp \! {\Bigl(i \mbf{q}_\perp \! \cdot \sum_{j=1}^{n-1} x_j \mbf{b}_{\perp j}\Bigr)}
 \left\vert  \psi_n(x_j, \mbf{b}_{\perp j})\right\vert^2
\end{equation}
where the normalization is defined by
\vspace{-4pt}
\begin{equation}  \label{eq:Normb}
\sum_n  \prod_{j=1}^{n-1} \int d x_j d^2 \mathbf{b}_{\perp j}
\vert \psi_{n/H}(x_j, \mathbf{b}_{\perp j})\vert^2 = 1.
\end{equation}

The formula  is exact if the sum is over all Fock states $n$.
For definiteness we shall consider a two-quark $\pi^+$  valence Fock state
$\vert u \bar d\rangle$ with charges $e_u = \frac{2}{3}$ and $e_{\bar d} = \frac{1}{3}$.
For $n=2$, there are two terms which contribute to the $q$-sum in (\ref{eq:FFb}).
Exchanging $x \leftrightarrow 1 \! - \! x$ in the second integral  we find
\begin{equation}  \label{eq:PiFFb}
 F_{\pi^+}(q^2)  =  2 \pi \int_0^1 \! \frac{dx}{x(1-x)}  \int \zeta d \zeta
J_0 \! \left(\! \zeta q \sqrt{\frac{1-x}{x}}\right)
\left\vert \psi_{u \bar d/ \pi}\!(x,\zeta)\right\vert^2,
\end{equation}
where $\zeta^2 =  x(1  -  x) \mathbf{b}_\perp^2$ and $F_{\pi^+}(q\!=\!0)=1$.

We now compare this result with the electromagnetic form-factor in AdS space:~\cite{Polchinski:2002jw}
\begin{equation}
F(Q^2) = R^3 \int \frac{dz}{z^3} \, J(Q^2, z) \vert \Phi(z) \vert^2,
\label{eq:FFAdS}
\end{equation}
where $J(Q^2, z) = z Q K_1(z Q)$.
Using the integral representation of $J(Q^2,z)$
\begin{equation} \label{eq:intJ}
J(Q^2, z) = \int_0^1 \! dx \, J_0\negthinspace \left(\negthinspace\zeta Q
\sqrt{\frac{1-x}{x}}\right) ,
\end{equation} we write the AdS electromagnetic form-factor as
\begin{equation}
F(Q^2)  =    R^3 \! \int_0^1 \! dx  \! \int \frac{dz}{z^3} \,
J_0\!\left(\!z Q\sqrt{\frac{1-x}{x}}\right) \left \vert\Phi(z) \right\vert^2 .
\label{eq:AdSFx}
\end{equation}
Comparing with the light-front QCD  form factor (\ref{eq:PiFFb}) for arbitrary  values of $Q$~\cite{Brodsky:2006uqa}
\begin{equation} \label{eq:Phipsi}
\vert \psi(x,\zeta)\vert^2 =
\frac{R^3}{2 \pi} \, x(1-x)
\frac{\vert \Phi(\zeta)\vert^2}{\zeta^4},
\end{equation}
where we identify the transverse LF variable $\zeta$, $0 \leq \zeta \leq \Lambda_{\rm QCD}$,
with the holographic variable $z$.
Identical results are obtained from the mapping of the QCD gravitational form factor
with the expression for the hadronic gravitational form factor in AdS space.~\cite{Brodsky:2008pf,Abidin:2008ku}

\section {Vacuum Effects and Light-Front Quantization}

The LF vacuum is remarkably simple in light-cone quantization because of the restriction $k^+ \ge 0.$   For example in QED,  vacuum graphs such as $e^+ e^- \gamma $  associated with the zero-point energy do not arise. In the Higgs theory, the usual Higgs vacuum expectation value is replaced with a $k^+=0$ zero mode~\cite{Srivastava:2002mw}; however, the resulting phenomenology is identical to the standard analysis.

Hadronic condensates play an important role in quantum chromodynamics (QCD).
Conventionally, these condensates are considered to be properties
of the QCD vacuum and hence to be constant throughout spacetime.
A new perspective on the nature of QCD
condensates $\langle \bar q q \rangle$ and $\langle
G_{\mu\nu}G^{\mu\nu}\rangle$, particularly where they have spatial and temporal
support,
has recently been presented.~\cite{Brodsky:2008be,Brodsky:2009zd,Brodsky:2008xm,Brodsky:2008xu}
Their spatial support is restricted to the interior
of hadrons, since these condensates arise due to the interactions of quarks and
gluons which are confined within hadrons. For example, consider a meson consisting of a light quark $q$ bound to a heavy
antiquark, such as a $B$ meson.  One can analyze the propagation of the light
$q$ in the background field of the heavy $\bar b$ quark.  Solving the
Dyson-Schwinger equation for the light quark one obtains a nonzero dynamical
mass and, via the connection mentioned above, hence a nonzero value of the
condensate $\langle \bar q q \rangle$.  But this is not a true vacuum
expectation value; instead, it is the matrix element of the operator $\bar q q$
in the background field of the $\bar b$ quark.  The change in the (dynamical)
mass of the light quark in this bound state is somewhat reminiscent of the
energy shift of an electron in the Lamb shift, in that both are consequences of
the fermion being in a bound state rather than propagating freely.
Similarly, it is important to use the equations of motion for confined quarks
and gluon fields when analyzing current correlators in QCD, not free
propagators, as has often been done in traditional analyses of operator
products. Since the distance between the
quark and antiquark cannot become arbitrarily large, one cannot create a quark
condensate which has uniform extent throughout the universe.  The $45$ orders of magnitude conflict of QCD with the observed value of the cosmological condensate is thus removed.~\cite{Brodsky:2008xu}
A new perspective on the nature of quark and gluon condensates in
quantum chromodynamics is thus obtained:~\cite{Brodsky:2008be,Brodsky:2008xm,Brodsky:2008xu}  the spatial support of QCD condensates
is restricted to the interior of hadrons, since they arise due to the
interactions of confined quarks and gluons.  In  LF theory, the condensate physics is replaced by the dynamics of higher non-valence Fock states as shown by Casher and Susskind.~\cite{Casher:1974xd}  In particular, chiral symmetry is broken in a limited domain of size $1/ m_\pi$,  in analogy to the limited physical extent of superconductor phases.  This novel description  of chiral symmetry breaking  in terms of ``in-hadron condensates"  has also been observed in Bethe-Salpeter studies.~\cite{Maris:1997hd,Maris:1997tm}
This picture also explains the
results of recent studies~\cite{Ioffe:2002be,Davier:2007ym,Davier:2008sk} which find no significant signal for the vacuum gluon
condensate.

AdS/QCD also provides  a description of chiral symmetry breaking by
using the propagation of a scalar field $X(z)$
to represent the dynamical running quark mass. The AdS
solution has the form~\cite{Erlich:2005qh,DaRold:2005zs} $X(z) = a_1 z+ a_2 z^3$, where $a_1$ is
proportional to the current-quark mass. The coefficient $a_2$ scales as
$\Lambda^3_{QCD}$ and is the analog of $\langle \bar q q \rangle$; however,
since the quark is a color nonsinglet, the propagation of $X(z),$ and thus the
domain of the quark condensate, is limited to the region of color confinement.
Furthermore the effect of the $a_2$ term
varies within the hadron, as characteristic of an in-hadron condensate. 
A similar solution is found in the soft wall model in presence of a positive sign dilaton.~\cite{Zuo:2009dz}
The AdS/QCD picture of condensates with spatial support restricted to hadrons
is also in general agreement with results from chiral bag 
models~\cite{Chodos:1975ix,Brown:1979ui,Hosaka:1996ee},
which modify the original MIT bag by coupling a pion field to the surface of
the bag in a chirally invariant manner.

\section{Conclusions}

We have derived a connection between a semiclassical first approximation to QCD, quantized on the light-front,
and hadronic modes propagating on a fixed AdS background. This
leads to an effective relativistic Schr\"odinger-like equation in the AdS fifth dimension coordinate $z$ (\ref{eq:QCDLFWE}).
We have show how this identical AdS wave equation can be derived in physical space time as an effective equation for valence quarks in LF quantized theory, where one identifies the AdS fifth dimension coordinate $z$ with the LF coordinate $\zeta$.
We originally derived this correspondence using the identity between electromagnetic and gravitational form factors computed in AdS and LF theory~\cite{Brodsky:2006uqa,Brodsky:2007hb,Brodsky:2008pf}. Our derivation shows that the 
fifth-dimensional mass $\mu$ in the AdS equation of motion
is directly related to orbital angular momentum $L$ in physical space-time. The result is physically compelling and phenomenologically successful.

We have shown how the soft-wall AdS/CFT model with a dilaton-modified AdS space leads to the
 potential $U(z) = \kappa^4 z^2 + 2 \kappa^2(L+S-1)$. This potential can be derived directly from the action in AdS space and corresponds to a dilaton 
profile  $\exp{(+\kappa^2 z^2)}$, 
 with  opposite sign to that of Ref.  \cite{Karch:2006pv}.  Hadrons are identified by matching the power behavior of the hadronic amplitude at the AdS boundary at small $z$ to the twist of its interpolating operator at short distances $x^2 \to 0$, as required by the AdS/CFT dictionary. The twist corresponds to the dimension of fields appearing in chiral super-multiplets. The twist of a hadron equals the number of constituents.

The Schr\"odinger-like light-front  AdS/QCD equation provides successful predictions for the light-quark meson and baryon spectra, as function of hadron spin, quark angular momentum, and radial quantum number.~\cite{note1} The pion is massless for zero mass quarks in agreement with chiral invariance arguments.
The predictions for form factors are also successful. The predicted power law fall-off agrees with dimensional counting rules as required by conformal invariance at small $z$.~\cite{Brodsky:2007hb,Brodsky:2008pg}

Higher spin modes follow from shifting dimensions in the AdS wave equations.
In the hard-wall model the dependence is linear:  $\mathcal{M} \sim 2n + L$.
In the soft-wall
model the usual Regge behavior is found $\mathcal{M}^2 \sim n +
L$.   Both models predict the same multiplicity of states for mesons
and baryons as observed experimentally.~\cite{Klempt:2007cp}
It is possible to extend the model to hadrons with heavy quark constituents 
by introducing nonzero quark masses and short-range Coulomb
corrections. 

The AdS/QCD semiclassical approximation to light-front QCD
does not account for particle
creation and absorption, and thus it will break down at short distances
where hard gluon exchange and quantum corrections become important. 
However, the model 
can be systematically improved  by using its complete orthonormal solutions to diagonalize the QCD light-front Hamiltonian~\cite{Vary:2009gt}  or by applying the Lippmann-Schwinger method to systematically include the QCD interaction terms.

\section*{Acknowledgments}
 Presented by SJB  at  the 10th Workshop on Non-Perturbative QCD at the
 Institut d'Astophysique de Paris (IAP), June 8-12, 2009.
 We thank Carl Carlson,  Alexandre Deur, Josh Erlich, Stan Glazek, Paul Hoyer,  Eberhard Klempt, Craig Roberts, Robert Shrock,  James Vary, and Fen Zuo  for helpful conversations and collaborations. 
Section 3 is based on collaborations with Robert Shrock.
This research was supported by the Department
of Energy  contract DE--AC02--76SF00515.  SLAC-PUB-13790.

\end{document}